%% file: neutral_paper.tex
\documentclass[10pt, a4paper, twoside]{amsart}

\input{style_PM}
\allowdisplaybreaks
\usepackage{amsfonts}
\usepackage{amsmath}
\usepackage{bbm}
\usepackage{float}
\usepackage{tikz}
\usepackage{tikz-3dplot}
\usepackage{hhline}
\usepackage{caption}
\usepackage{subcaption}
\usetikzlibrary{arrows,chains,matrix,positioning,scopes}
\usetikzlibrary{shapes,calc,positioning}

\makeatletter
\tikzset{join/.code=\tikzset{after node path={%
\ifx\tikzchainprevious\pgfutil@empty\else(\tikzchainprevious)%
edge[every join]#1(\tikzchaincurrent)\fi}}}
\makeatother
\tikzset{>=stealth',every on chain/.append style={join},
         every join/.style={->}}
\tikzstyle{labeled}=[execute at begin node=$\scriptstyle,
   execute at end node=$]

\newcommand\ddfrac[2]{\frac{\displaystyle #1}{\displaystyle #2}}

\title[Distance Based Source Domain Selection for Sentiment Classification]
{Distance Based Source Domain Selection for Sentiment Classification}

\author[L.E. Razoux Schultz, M. Loog and P. Mohajerin Esfahani]{Lex Elias Razoux Schultz, Marco Loog, Peyman Mohajerin Esfahani}%
	\thanks{The authors are with the Delft Center for Systems and Control, TU Delft, The Netherlands and the Patter Recognition Laboratory, TU Delft, The Netherlands}

\begin{document}
\maketitle

\begin{abstract}
    Automated sentiment classification (SC) on short text fragments has received increasing attention in recent years. Performing SC on unseen domains with few or no labeled samples can significantly affect the classification performance due to different expression of sentiment in source and target domain. In this study, we aim to mitigate this undesired impact by proposing a methodology based on a predictive measure, which allows us to select an optimal source domain from a set of candidates. The proposed measure is a linear combination of well-known distance functions between probability distributions supported on the source and target domains (e.g. Earth Mover's distance and Kullback-Leibler divergence). The performance of the proposed methodology is validated through an SC case study in which our numerical experiments suggest a significant improvement in the cross domain classification error in comparison with a random selected source domain for both a naive and adaptive learning setting. In the case of more heterogeneous datasets, the predictability feature of the proposed model can be utilized to further select a subset of candidate domains, where the corresponding classifier outperforms the one trained on all available source domains. This observation reinforces a hypothesis that our proposed model may also be deployed as a means to filter out redundant information during a training phase of SC.
\end{abstract}

\section{Introduction} \enlargethispage{\baselineskip}

Automated SC is performed when a trained classifier is used to label documents with a sentiment label, based on the content of the document. In practice, a document represents a review, tweet or other small textual opinion expression and ought to be classified as either positive or negative (binary SC). Allied terminology for SC is semantic orientation or polarity, which indicates the direction a word deviates from the norm of its semantic group \cite{hatzivassiloglou1997predicting}. Often, the terms opinion mining and sentiment analysis are used to describe the computational treatment of opinion, sentiment and subjectivity in text \cite{pang2008opinion}. Examples of current applications that benefit from SC are recommendation systems \cite{terveen1997phoaks}, stock market prediction systems \cite{bollen2011twitter} and political election predictors \cite{tumasjan2010predicting}. Also in the future, we could imagine SC put in practice to improve man-machine communication such as voice commands or even interaction between robots and humans.

One of the biggest challenges in many SC applications, is the discrepancy between the source domain where the classifier is trained on, and the target domain of interest. Typically, we have large amounts of labeled data from different source domains, but only possess few, unlabeled data from the target domain, both originated from different underlying distributions. In this setting, SC generally performs rather bad compared to inner domain training and testing. This so called domain transfer problem can be intuitively explained due to different fashions of sentiment expression in different domains \cite{tan2009adapting}. A word may have different sentiment polarity with respect to the domains it is used in, or different words may be used to express similar sentiment among different domains \cite{zhang2015cross}. 

Various research has been performed with the aim to decrease the loss in classification performance due to domain shift. This loss is referred to as the adaptation loss \cite{he2011automatically} \cite{blitzer2007biographies}. One of the popular transfer learning techniques to maintain performance of SC when crossing domains, is Structural Correspondence Learning which uses pivot features as link between the source and target domain \cite{blitzer2006domain} \cite{kouw2016feature}. Many other techniques are based on optimal transport which matches the conditional probability distribution of the training domain and target domain by a transformation of the source domain data  \cite{courty2015optimal} \cite{kolouri2016continuous}. The distance between source and target distribution is minimized \cite{cohen1999earth}. Often distance is minimized in a lower dimensional embedded space, for example when performing Subspace Alignment \cite{fernando2013unsupervised} or Transfer Component Analysis \cite{pan2011domain}. 

In recent years, the use of neural models for adaptation learning has been proven successful. The ability of a neural model to be trained on both source and unlabeled target data has been leveraged to improve performance of the cross domain classification task. Various approaches are proposed to adapt the neural net towards superior performance of classification in a given target domain. In the field of SC, Domain-Adversarial Neural Networks (DANNs) have been effectively used in such a context \cite{ganin2016domain}. DANNs improve performance by augmenting a feed-forward model with a few standard layers and a new gradient reversal layer. Due to this augmentation, the model promotes features that are discriminative on class label and indiscriminative regarding a shift in domain. Other research is based on the use of Stacked Denoising Auto-encoders (SDA's) \cite{chen2012marginalized} \cite{glorot2011domain}. An SDA can be used to learn a higher lever feature extraction by using unlabeled data from the source and target domain. The feature extraction transforms the feature space, and subsequently, a classifier is trained on this transformed feature space. Other popular techniques with much resemblance to SDA's are the \textit{word2vec} \cite{mikolov2013efficient} and \textit{doc2vec} \cite{le2014distributed} algorithms.

One may invoke other techniques in combination with the adaptation learning to improve the adaption loss, e.g., unsupervised source domain selection.  Instead of focusing on the type of classifier or the underlying representation of the data, these techniques primarily revolve around preselecting an appropriate training dataset. The source domain selection method opts to select one or multiple optimal source domains from a set of candidate domains without supervision, i.e., it selects the source domains that have the best cross domain classification performance compared to the other candidates without the use of class labels of the target domain data. Source domain selection has been successfully deployed in other contexts, for example to improve brain-computer interface calibration \cite{wu2015reducing}. In this case, supervised source domain selection based on distance between the class average vectors has been used. Alternatively, ranking convolutional neural networks trained on difference source domains for a target task shows success when using the mutual information between domains \cite{afridi2018automated}.

Source domain selection is a higher level of selection than instance selection as it chooses a set of instances from the same underlying domain rather than individual instances. It is based on the assumption that inner domain information is more informative and therefore yields an improved performance when used for training. That rises the question what constitutes a domain. It is argued that there is no common ground in what constitutes a domain, but in NLP it is typically used to refer tot some coherent data with respect to topic or genre \cite{plank2016non}. However, there are numerous other factors that should be taken into consideration, e.g., the medium on which the domain data is exposed, the time spirit, and the purpose of the data.

In order to find an appropriate source domain, it is recently proposed to represent a domain as an average of its document representations in a pre-trained embedded space \cite{ruder2017data}. Additionally, this study performs same experiments when using denoising auto-encoder representations trained on all source data instead of a pre-trained embedding. They conclude that using a pre-trained embedded space for domain representation is not necessarily recommended. Their approach that relies on denoising auto-encoders is showed to be effective in selecting source domains, but only when large training sets are available.

More common approaches for source domain selection use domain similarity metrics. One of these metrics developed for assessing potential of transfer learning is the $\mathcal{A}$-distance \cite{ben2007analysis}. In the field of source domain selection for SC, Blitzer et al. show a good correlation between the $\mathcal{A}$-distance and adaptation loss  \cite{blitzer2007biographies}. The adaptation loss is measured using the target domain's inner domain classification error which unfortunately requires labeled data of the target domain. The concept of $\mathcal{A}$-distance is also used in recent research for orthophoto classification, were it is approximated by the Maximum Mean Discrepancy \cite{vogt2017boosted}. For a more open search space for a source domain, techniques have been researched to find a suitable source domain in open online information sources such as Wikipedia \cite{xiang2011source}. However, this technique also uses labeled information of the target domain. 

Another similarity metric used in source domain selection is the Kullback-Leibler divergence (KL) \cite{kullback1951information}. This metric may not be well defined in SC, and therefore the data requires smoothing. Alternatively, approximations to KL, such as the Jensen-Shannon divergence \cite{lin1991divergence} or the skew divergence \cite{lee2001effectiveness} can be used to determine distance between domains. These divergences indicators have proven their ability to select a suitable source domain and thus improving the classification performance of cross domain classification tasks \cite{plank2011effective}. 

In this article, we propose a new method, CMEK, for source domain selection that does not require any labeled information of the target domain. The CMEK method trains a source domain selection model by means of a set of given distance metrics. This approach differs from other state-of-the art approaches as it proposes a linear combination of distance metrics where weights are determined by training on the available source domain instead of a single metric for source domain selection. We emphasize that the proposed technique here can be utilized in combination with adaptation learning in a sequential manner, which potentially improves the performance of our domain transformation task.

In Section \ref{problemdef}, we define the problem of source domain selection in general, that is, not in context of SC. Section \ref{method} proposes an approach to predict the performance of a classifier trained on a source domain, and applies it on a target domain of interest of which the labels are unknown. This prediction is based on statistical distances between the distributions of the source and target data. To this end, we leverage several statistical measures to determine the distance. Section \ref{limitations} addresses certain limitations to our proposed model, starting with fundamental limitations that occur in all classification tasks and ending with limitations of our approach within the task of SC. In Section \ref{case_study} we present background knowledge on how to perform sentiment classification by using a bag-of-words model, as we do in our experiments. Section \ref{ex_setup} describes the experimental setup and model settings for two different types of datasets; a homogeneous one where the datasets share similar contents and a heterogeneous one that contains datasets from very diverse topics and sources. In this article, our homogeneous set used is the DRANZIERA benchmark dataset \cite{dragoni2016dranziera}. In addition, we describe how we run experiments in an adaptation learning setting. Section \ref{results} reports the numerical results in detail, and finally in Section \ref{conclusion} we conclude with our reflection on the results and final messages along with suggestions for future work.

\section{Problem Definition} \label{problemdef}
With a document represented by a random variable $X$ and its sentiment label by random variable $Y$, each domain is uniquely characterized by its joint probability density function supported on $\mathbb{X} \times \mathbb{Y}$. Let us then define two underlying distributions to the evaluated source and target data on $\mathbb{X} \times \mathbb{Y}$, denoted by $\mathbb{P}$ and $\bar{\mathbb{P}}$ respectively. Their marginals on $\mathbb{X}$, are denoted as $\mathbb{P}_x$ and $\bar{\mathbb{P}}_x$ respectively. In machine learning, a realization of $\mathbb{P}$ is referred to as labeled data whereas realizations from $\mathbb{P}_x$ are referred to as unlabeled data.

We define an hypothesis function as a mapping from $\mathbb{X}$ to $\mathbb{Y}$, $h:\mathbb{X} \rightarrow \mathbb{Y}$. In the context of classification problems in the machine learning literature, the set $\mathbb{Y}$ is often discrete, often binary, and the hypothesis $h$ is often referred to as a classifier as it is able to assign a class label in $\mathbb{Y}$ to each data point in $\mathbb{X}$. The true hypothesis function $h^{\star}_{\mathbb{P}}$ for a domain characterized by $\mathbb{P}$, is defined as

\begin{equation}
\label{eqn:eqhstar}
h^{\star}_{\mathbb{P}}:= \arg\min_{h} {\rm Prob}\big(h(x) \neq y ~|~ (x,y) \sim \mathbb{P} \big).
\end{equation}

We then define the inner domain classification error as the probability of misclassification in the same domain as the true hypothesis function is constructed in. And the cross domain classification error is defined as the probability of misclassification in the target domain of the true hypothesis function of the source domain

\begin{align}
\label{eqn:crossdomainerror}
\begin{split}
\xi(\mathbb{P}) &:= {\rm Prob} \big(h^{\star}_{\mathbb{P}}(x) \neq y ~|~ (x,y) \sim \mathbb{P}\big),
\\
\xi(\mathbb{P},\bar{\mathbb{P}}) &:= {\rm Prob}\big(h^{\star}_{\mathbb{P}}(x) \neq y ~|~ (x,y) \sim \bar{\mathbb{P}}\big).
\end{split}
\end{align}

\noindent Note that due to the stochastic nature of sentiment expression, $\xi(\mathbb{P})$ and $\xi(\mathbb{P},\bar{\mathbb{P}})$ are not likely to be zero, even if we would have infinite data to train on. In addition, due to the difference in source and target distribution, $\xi(\mathbb{P},\bar{\mathbb{P}})$ is likely to be higher than $\xi(\mathbb{P})$ \cite{blitzer2007biographies}. The difference between the cross and inner domain classification error is referred to as the adaptation loss.

The goal in many classification tasks is to minimize the cross domain classification error \eqref{eqn:crossdomainerror} for a specific target domain while we have access to a finite set of labeled candidate source domains. 

An appropriate choice of the source domain in this context is the main goal of this study, leading to the following question:

\begin{center}
\emph{What source domain minimizes the cross domain classification error for a given target domain?} 
\end{center}

With the choice of source domain restricted to a domain characterized by a probability density function in the candidate set $\mathcal{P}$, our main goal can be formally described through the optimization program

\begin{equation}
\label{eqn:objective}
\mathbb{P}^\star:= \arg\min_{\mathbb{P}\in\mathcal{P}} \xi(\mathbb{P},\bar{\mathbb{P}}).
\end{equation}

\noindent where $\xi(\mathbb{P},\bar{\mathbb{P}})$ is the cross domain classification error introduced in \eqref{eqn:crossdomainerror}. In other words, when we posses a finite number of labelled datasets, each originated from a unique source domain, on which set should we train our classifier in order to minimize classification error on data from a specific target domain?

The challenge concerning the objective \eqref{eqn:objective}, is that for a target domain of interest we typically only have unlabeled data, therefore we cannot calculate the cross domain classification error. This means we only know the marginal distribution $\bar{\mathbb{P}}_x$ of our target domain instead of $\bar{\mathbb{P}}$, making it impossible to explicitly calculate $\xi(\mathbb{P},\bar{\mathbb{P}})$. In order to find the best source domain characterized by $\mathbb{P}^\star \in \mathcal{P}$ for our target domain with distribution $\bar{\mathbb{P}}$, we ought to predict $\xi(\mathbb{P},\bar{\mathbb{P}})$ for all $\mathbb{P} \in \mathcal{P}$. For this prediction we have full distributional information of the source domains, $\mathbb{P}$, and marginal information on the distribution of the target domain, $\bar{\mathbb{P}}_x$, available.


\section{Method} \label{method}
In this section, we propose a model, called CMEK source domain selection, an acronym of the four measures we chose in our case study, to deal with the previously stated challenge. The model measures statistical distances between the marginal distributions of candidate source domains and the target domain and uses the inner domain classification error of the candidate source domains. The candidate source domain with the lowest distance is hypothesized to have the lowest cross domain classification error and is selected to train a classifier. This section describes the methodology, the distance measures we consider and how we evaluate the performance of the model.

We consider a set $\mathcal{D}$ containing a family of distance functions $d: \mathbb{P} \times \bar{\mathbb{P}}_x \rightarrow \mathbb{R}_+$. We now hypothesize that given an available set of source domain distributions $\mathcal{P}$ and a target domain distribution denoted by $\bar{\mathbb{P}}$ with the marginal $\bar{\mathbb{P}}_x$, there exists a $\hat{d} \in \mathcal{D}$ that can reliably predict the cross domain classification error $\xi(\mathbb{P},\bar{\mathbb{P}})$, that is,

\begin{equation}
\label{eqn:hypothesis}
\xi(\mathbb{P},\bar{\mathbb{P}}) \stackrel{\text{hyp}}{\approx} \hat{d}(\mathbb{P},\bar{\mathbb{P}}_x).
\end{equation}

More specific, we hypothesize that the cross domain classification error can be predicted by looking at the difference in marginal distribution functions $\mathbb{P}_x$ and $\bar{\mathbb{P}}_x$, and $\xi(\mathbb{P})$. With this prediction, the best candidate source domain can be selected for training. We formalize this optimal choice of the measure by considering the optimization problem

\begin{equation}
\label{eqn:optimal_d}
\hat{d}:= \arg \min_{d \in \mathcal{D}} \sum\limits_{\mathbb{P} \in \bar{\mathcal{P}}} \big|\xi(\mathbb{P},\bar{\mathbb{P}}) - d(\mathbb{P},\mathbb{P}_x)\big|.
\end{equation}

To construct the family $\mathcal{D}$ of candidate measures, we use a vector $s$ with $K$ known statistical distance measures as elements, $s_i: \mathbb{P}_x \times \bar{\mathbb{P}}_x \rightarrow \mathbb{R}_+$ for $i \in \{1,\cdots,K\}$. The family $\mathcal{D}$ consists of linear combinations of these measures $s_i$ with corresponding weight coefficients $\beta_i \in \mathbb{R}_+$, with $\beta_i$ being elements of the weight vector $\beta$. We augment the linear combination with a constant. Note that the distance measures are functions supported on the marginal distributions of source and target domain. Examples of these statistical distance metrics are Integral Probability Metrics and $\phi$-divergence. When a source domain has poor inner domain classification performance, we expect it to be less suitable as source domain for cross domain classification. Therefore, we calculate the inner domain classification error for each candidate source domain and add this to our vector $s$. With our choice for $\mathcal{D}$, the optimization problem \eqref{eqn:optimal_d} can be reduced to 

\begin{align}
\label{eqn:optimal_ds}
\begin{split}
\hat{\beta} &:= \arg \min_{\beta \geq 0} \sum\limits_{\mathbb{P} \in \mathcal{P}} \big|\xi(\mathbb{P},\bar{\mathbb{P}}) - \beta s(\mathbb{P},\bar{\mathbb{P}}_x)\big|,
\\
\hat{d} &:= \hat{\beta} s(\mathbb{P},\bar{\mathbb{P}}_x).
\end{split}
\end{align}

\noindent Note that program \eqref{eqn:optimal_ds} constructs the optimal unsupervised predictor by using the full distribution of the target data, which is presumed to be unknown. In a practical setting with a finite number of elements in $\mathcal{P}$, we deal with this problem by training only on the domains in $\mathcal{P}$. In each run, one of the elements of $\mathcal{P}$ is extracted from the set and used as proxy $\bar{\mathbb{P}}$, at the end of the run, this element, denoted as $\tilde{\mathbb{P}}$ and its marginal $\tilde{\mathbb{P}}_x$, is placed back in $\mathcal{P}$. In every run, another element is extracted and placed back for training. This training program is described as

\begin{equation*}
\hat{\beta} := \arg \min_{\beta \geq 0} 
\sum\limits_{\mathbb{P} \in \mathcal{P},\mathbb{P} \neq \tilde{\mathbb{P}}} 
\sum\limits_{\tilde{\mathbb{P}} \in \mathcal{P}}
\big|\xi (\mathbb{P},\tilde{\mathbb{P}}) - \beta s(\mathbb{P},\tilde{\mathbb{P}}_x)\big|.
\end{equation*}

We hypothesize that with a sufficient amount of source domain distributions in $\mathcal{P}$, the constructed predictor for the cross domain classification error, based on information from $\mathcal{P}$, is also reasonably accurate in predicting the cross domain classification error for a target domain not included in $\mathcal{P}$. This allows us to select the best source domain to train on for this unseen target domain.

Now, in view of our hypothesis \eqref{eqn:hypothesis}, with the predictor $\hat{d}$ we are able to approximate $\xi(\mathbb{P},\bar{\mathbb{P}})$ with only marginal information of the target domain $\bar{\mathbb{P}}_x$, which allows us to deal with the challenge through an approximation. This enables us to, given a target domain with distribution $\bar{\mathbb{P}}$, select the source domain, characterized by $\hat{\mathbb{P}}$ from a set of available source domains that is predicted to minimize the cross domain classification error $\xi(\mathbb{P},\bar{\mathbb{P}})$ as formalized through the optimization program

\begin{equation*}
\hat{\mathbb{P}} := \arg \min_{\mathbb{P} \in \mathcal{P}} \hat{d}(\mathbb{P},\bar{\mathbb{P}}_x)
\end{equation*}

\noindent which is the goal of this study.

\subsection{Measures}
In the SC case study, we hypothesize that the way people express sentiment can be captured in how often certain words and word combinations are used. The underlying joint distributions $\mathbb{P}$ differ among different domains. To accurately classify sentiment for a target domain, we therefore ought to find a domain with a similar underlying distribution function. To this end, one may need to measure the similarity between different distributions. In this article we consider two classes of distance measures: Integral Probability Metrics (IPM's), and $\phi$-divergences that map $\mathbb{X} \times \mathbb{X} \rightarrow \mathbb{R}$. Both measures compare the distributions of the features but in different fashions.

\subsubsection{Integral Probability Metrics}
The main characteristic component of IPM distance functions is the set $\mathcal{F}$ that contains a family of functions $f:\mathbb{X} \rightarrow \mathbb{R}$. Using $\mathcal{F}$, we define the distance between two distributions $\mathbb{P}_x$ and $\bar{\mathbb{P}}_x$ as 

\begin{equation*}
    \label{eqn:ipm}
    s_i(\mathbb{P}_x,\bar{\mathbb{P}}_x) := \sup_{f \in \mathcal{F}}\Big| \mathbb{E}^{\mathbb{P}_x}\big[f(X) \big] - \mathbb{E}^{\bar{\mathbb{P}}_x}\big[f(X) \big]   \Big|.
\end{equation*}

Different choices of the family $\mathcal{F}$ give rise to different distance functions. A popular choice is to limit our search space to the unit ball in a predefined metric space. In this article, we consider two choices: (i) the unit ball in a Reproducing Kernel Hilbert Space which corresponds to the Maximum Mean Discrepancy (MMD) distance function \cite{gretton2012kernel}, (ii) the unit ball of Lipschitz functions which corresponds to the Earth Mover's Distance (EMD) distance function \cite{arjovsky2017wasserstein}.

More intuitively, the EMD metric measures the minimum transportation cost to transform one distribution to another. The transportation cost, roughly speaking, amounts to the effort measured in a so-called ground metric that is required to move a unit probability mass. The MMD metric finds a well behaved, smooth, function which is typically high on the points drawn from $\mathbb{P}_x$ and low on $\bar{\mathbb{P}}_x$ or vice verse.

\subsubsection{Phi-divergence}
For the second class of distance measures, we consider the $\phi$-divergence between two distributions $\mathbb{P}$ and $\bar{\mathbb{P}}$ defined as 

\begin{equation*}
    \label{phid}
    s_i( \mathbb{P}_x, \bar{\mathbb{P}}_x ) :=\begin{cases}
    \int\limits_{\mathbb{X}} \phi \left( \ddfrac{d\mathbb{P}_x}{d\bar{\mathbb{P}}_x} \right) d\bar{\mathbb{P}_x}  & \mathbb{P}_x \ll \bar{\mathbb{P}}_x, \\
    + \infty,       & \text{otherwise}
    \end{cases}
\end{equation*}

\noindent where $\phi$ is a convex function, and $\mathbb{P}_x \ll \bar{\mathbb{P}}_x$ denotes that $\mathbb{P}_x$ is absolutely continuous with respect to $\bar{\mathbb{P}}_x$ \cite{sriperumbudur2009integral}. A widely used choice is $\phi(t) = t \log (t)$, resulting in the Kullback-Leibler Divergence (KLD). Another popular choice for $\phi(t)$ is $\phi(t) = (t-1)^2$ which is based on Pearson's $\chi^2$ test statistic. This choice gives the $\chi^2$-divergence (Chi2) distance measure. An interesting property of these two divergences is that they are asymmetric, $s_i(\mathbb{P}_x,\bar{\mathbb{P}}_x) \neq s_i(\bar{\mathbb{P}}_x,\mathbb{P}_x)$. 

Since we hypothesize that the way people express sentiment can be captured in how often certain words and word combinations are used, regardless of the context, we are then left with a broad spectrum of possibilities to optimize over. We propose to use a linear combination of the two IPM approaches MMD and EMD, the two $\phi$-divergences KLD and Chi2, a constant representing an offset error and the inner domain classification error of the source domain. 

In most practical cases, we only have access to the true probability distribution through a finite number of observations. For that reason, in order to evaluate these discrete distributions, the integrals are replaced by summation. Selection based on the linear combination of these four metrics, the constant, and $\xi(\mathbb{P})$ is referred to as the CMEK selection model.

\begin{equation*}
\begin{split}
    d(\mathbb{P},\bar{\mathbb{P}}_x) = &\beta_1 Chi2(\mathbb{P}_x,\bar{\mathbb{P}}_x) +\beta_2 MMD(\mathbb{P}_x,\bar{\mathbb{P}}_x) \\+ &\beta_3 EMD(\mathbb{P}_x,\bar{\mathbb{P}}_x) + \beta_4 KLD(\mathbb{P}_x,\bar{\mathbb{P}}_x)\\ + &\beta_5 \xi(\mathbb{P}) + \beta_0
\end{split}
\end{equation*}

\subsection{Performance evaluation} \label{evaluation}
We have constructed a model that predicts cross domain classification error between a candidate source domain and a target domain with a linear combination of statistical distance measures based on the marginal distributions $\mathbb{P}_x$ and $\bar{\mathbb{P}}_x$, a constant and $\xi(\mathbb{P})$. Now we are interested in how well this model is able to identify the best candidate source domain, i.e. the one that gives the lowest cross domain classification error $\xi(\mathbb{P},\bar{\mathbb{P}})$. For $n_c$ acquired datasets, let us use one corpus as target domain, $\bar{\mathbb{P}}$, and the others as candidate source domains with the set annotation $\mathcal{P}^{\prime}$. We can repeat the experiment $n_c$ times by using each time a different target domain and average the results. 

We let the model select the predicted best source domain $\hat{\mathbb{P}} \in \mathcal{P}^{\prime}$ and use this domain for training. To evaluate performance we define the relative cross domain classification error as
\begin{equation}
\label{eqn:relative_error}
\xi_{relative}(\hat{\mathbb{P}},\bar{\mathbb{P}}) = \xi(\hat{\mathbb{P}},\bar{\mathbb{P}})-\xi(\mathbb{P}^{\star},\bar{\mathbb{P}})
\end{equation}

\noindent where $\mathbb{P}^{\star}$ is the domain in $\mathcal{P}^{\prime}$ of which we know it has the lowest cross domain classification error, the true best domain. When the model selects the best source domain, the relative error will therefore be zero. We compare the distribution of this relative error when using the CMEK model with randomly selecting one source domain for training. Also, we look at the absolute cross domain classification error, the probability of selecting the best domain and the probability of selecting one of the five worst domains when using the CMEK model. We compare this with randomly selecting a domain and with an optimal selection model. 

Next, we let our proposed CMEK model select the best $n$ domains from $\mathcal{P}^{\prime}$ and compare $\xi(\bigcup_{i=1}^{n} \hat{\mathbb{P}}_i,\bar{\mathbb{P}})$ with training on all domains in $\mathcal{P}^{\prime}$ and with randomly selecting $n$ domains. All results are tested on significance using a paired t-test over the results of 13 runs, we use a significance cutoff of $p=0.05$. The paired t-test is strongly dependent on the assumption that the pairwise differences are normally distributed. For every comparison between results, we evaluated the normality assumption. We report only on significant results for which the normality assumption is confirmed.

To test if our methodology creates added value when using it in an adaptive learning setting, we let our model select the predicted best source domain and evaluate the cross domain classification error when using a DANN adaptive learning setting described in the recent study \cite{ganin2016domain}. As a benchmark, the result is compared with random selecting a source domain in the same training and test setting.

\section{Limitations} \label{limitations}
In solving the problem as defined in section \ref{problemdef} with our proposed method, we encounter limitations of various kinds. We first discuss a fundamental statistical limitation that is present in almost all practical machine learning problems. Then we discuss a limitation of computational nature and how to deal with this. In the last paragraph we elaborate on limitations that are more specific to the case study of SC: numerically representing text and the limited hypothesis space. We end with briefly explaining the limitations of our proposed CMEK model.


\subsection{Statistical limitations}
In the first place, and inherent to all classification problems, we are unable to retrieve the true distribution over the support set $\mathbb{X} \times \mathbb{Y}$ in practical settings. Instead, we need to infer this distribution through only a finite number of observations. In this case a natural approximation for the distribution is the empirical distribution supported on these observations, which are the available documents in our SC case study. We then encounter an inevitable approximation

\begin{equation*}
\label{eqn:cont2disc}
\mathbb{P} \approx \frac{1}{n_s} \sum\limits_i^{n_{s}}\delta_{\{x_i,y_i\}}
\end{equation*}

\noindent where $n_s$ denotes the number of documents in the domain sample. Due to the approximation of the true distribution, the hypothesis function \eqref{eqn:eqhstar} is suboptimal, resulting in a higher classification error. To evaluate this error, the available data originated from one domain is split into a train and test set. The train set is used to construct $h^{\star}_{\mathbb{P}}$ as defined in \eqref{eqn:eqhstar}. The test set is used to calculate the inner domain classification error. When using sufficient data in the test set, a large difference in apparent and true error indicates that the observations used for training are a poor support for approximating the true distribution since it means that the empirical distributions from the train and test set are not much alike. When this is the case, we are over fitting; the hypothesis function works for the data it is constructed on, but can not be properly generalized to perform well on new data from the same underlying distribution. To reduce over fitting, we ought to use a sufficient number of training objects and a low but not too low number of features.


\subsection{Computational limitations}
Another limitation is that the optimization in \eqref{eqn:eqhstar} uses the indicator function as loss function, which is non-convex. For computational reasons, we replace this loss function with a convex counterpart. For the true hypothesis function $h_{\mathbb{P}}^{\star}$ in \eqref{eqn:eqhstar}, a common convex loss function $\ell(x,y)$ to minimize, is the quadratic error of the prediction, popular as the method of least squares

\begin{equation}
\label{eqn:hoptimal_loss}
h^{\star}_{\mathbb{P}}= \arg\min_{h} \sum\limits_{i=1}^{n_s} \ell(x_i,y_i) \qquad \ell(x,y):=\big( h(x)-y \big)^2,
\end{equation}

\noindent and the cross domain classification error $\xi(\mathbb{P},\bar{\mathbb{P}})$ is then efined by empirical distributions of the target and the hypothesis function from \eqref{eqn:hoptimal_loss},

\begin{equation}
\label{eqn:inner_domain_error}
\xi(\mathbb{P},\bar{\mathbb{P}}):= \frac{1}{n_s} \sum\limits_{i}^{n_s} \mathbbm{1}_{\big\{h^{\star}_{\mathbb{P}}(x_i) \neq y_i\ ~|~ (x,y)\sim \bar{\mathbb{P}} \big\}}.
\end{equation}


\subsection{Further discussion on SC related limitations}
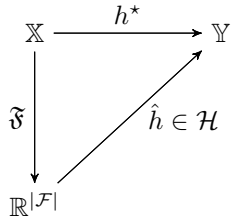
\begin{figure}[ht]
\centering
    \begin{tikzpicture}
      \matrix (m) [matrix of math nodes, row sep=5em, column sep=5em]
        { \mathbb{X} & \mathbb{Y}\\
          \mathbb{R}^{|\mathcal{F}|} & \\ };
          
      { [start chain] \chainin (m-1-1); \chainin (m-1-2) [join={node[above] {$h^{\star}$}}];}
    
      { [start chain] \chainin (m-1-1); \chainin (m-2-1) [join={node[left] {$\mathfrak{F}$}}];} 
      
      { [start chain] \chainin (m-2-1); \chainin (m-1-2) [join={node[right] {$~\hat{h} \in \mathcal{H}$}}];}
    \end{tikzpicture}
    \caption{Representation and classification structure} 
    \label{fig:classification_scheme}
\end{figure}

There are limitations concerned with performing SC with a nature that may also be relevant to other fields of application. Choosing a word representation model is a field of study by itself and should be considered carefully here. In the previous section we assumed a true hypothesis function $h^{\star}$ mapping each document represented in $\mathbb{X}$ to a label in $\mathbb{Y}$. For computational purpose, we need to numerically express documents in the $\mathbb{X}$ space. Let us call $W$ the set of all words, $W = \{ w_1, w_2, \dots w_{n_w} \}$ where $n_w$ is the number of unique words. Then, documents in $\mathbb{X}$ are combinations of these words, indicating that $\mathbb{X}$ can be viewed as an element in the power set of $W$ denoted by $\wp(W)$. Note that repetition in the subset is possible. For computational feasibility, we choose a subset $\mathcal{F} \subset \wp(W)$ to represent our documents. The subset $\mathcal{F}$ is called the feature set whose cardinality is denoted by $|\mathcal{F}|$. The features selection process depends on what representation model is at hand. The representation model $\mathfrak{F}$ maps $\mathbb{X}$ to $\mathbb{R}^{|\mathcal{F}|}$, i.e. $\mathfrak{F}:\mathbb{X} \rightarrow \mathbb{R}^{|\mathcal{F}|}$. The computational need for a representation model $\mathfrak{F}$ to select a subset of $\mathbb{X}$ for representation, limits the classification model since information has to be discarded, see Figure \ref{fig:classification_scheme}.

Furthermore, in the task of classification we need to define a search space $\mathcal{H}$ for our hypothesis function as it is computational impossible to search through all possible hypothesis functions as proposed in \eqref{eqn:eqhstar}. SC tasks are characterized by the fact that the representation space is sparse, i.e. the ratio of number of training documents to $|\mathcal{F}|$ is low. In this situation, computational feasibility calls for a tractable subset of hypothesis functions $\mathcal{H}$ such as all linear functions. The true hypothesis function is not likely to be in our hypothesis space, $h^{\star} \not\in \mathcal{H}$. We define the optimal hypothesis function constructed in the domain $\mathbb{P}$ as the hypothesis $\hat{h}$ within our limited hypothesis space $\mathcal{H}$, which is the collection of all linear mappings from $\mathbb{R}^{|\mathcal{F}|}$ to $\mathbb{R}$, that minimizes the chosen loss function \eqref{eqn:hoptimal_loss}. The cross domain classification error from \eqref{eqn:inner_domain_error} is calculated with the optimal hypothesis function $\hat{h}_{\mathbb{P}}$.

Hereafter, we use $\mathbb{P}$ to annotate the empirical distribution instead of the true continuous distribution. Furthermore, the cross domain classification error defined as $\xi(\mathbb{P},\bar{\mathbb{P}})$ in \eqref{eqn:crossdomainerror} is calculated with the optimal hypothesis function in $\mathcal{H}$ that minimizes the empirical loss in \eqref{eqn:hoptimal_loss}.


\subsection{CMEK selection model limitations}
Our model assumes that a good source domain, with low cross domain classification error, can be identified by the distributions $\mathbb{P}$ and $\bar{\mathbb{P}}_x$. However, where in reality sentiment is determined by joint distribution of the features, the measure we use only measure distance on the individual distributions of the features. This means we try to find similarity of the joint distributions from marginal distributions of individual features. The same holds for classification that is based on marginal data of individual features whereas sentiment in real life is determined by combinations of features. Since we use marginal data for calculating the cross domain classification error as well as the prediction of this error, this approach seems justified, but it creates a model separated from the true rules of expressing sentiment. 

In addition, we do not know in what way we should compare distributions, only that when distributions are the same, we expect no adaptation loss. It might very well be that the linear combination of our chosen measures does not include the true distance function that distinguishes domains from each other.


\section{Case Study: Sentiment Classification} \label{case_study}
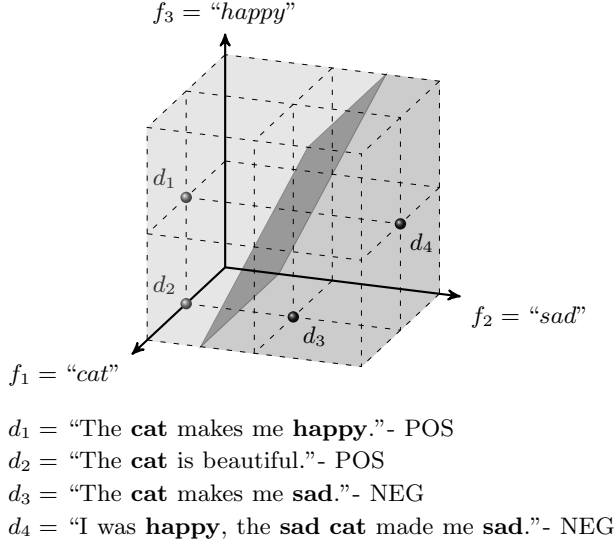
\begin{figure}[ht]
    \centering
    \tdplotsetmaincoords{70}{110}
    \begin{tikzpicture}[scale=3,tdplot_main_coords]
        
        \coordinate  (g1) at (0,0,0){};
        \coordinate  (g2) at (1,0,0){};
        \coordinate  (g3) at (1,0,1){};
        \coordinate  (g4) at (0,0,1){};
        \coordinate  (g5) at (0,1,0){};
        \coordinate  (g6) at (1,1,0){};
        \coordinate  (g7) at (1,1,1){};
        \coordinate  (g8) at (0,1,1){};
        \coordinate  (g9) at (0,0.25,0){};
        \coordinate  (g10) at (1,.25,0){};
        \coordinate  (g11) at (0,0.75,1){};
        \coordinate  (g12) at (1,.75,1){};
        \draw[fill=black, opacity = 0.1] (g2) -- (g10) -- (g12) -- (g11) -- (g4) -- (g3) -- cycle;
        \draw[fill=black, opacity = 0.2] (g10) -- (g6) -- (g5) -- (g8) -- (g11) -- (g9) -- cycle;
        \draw[fill=black, opacity = .4] (g9) -- (g10) -- (g12) -- (g11) -- cycle;

        \coordinate  (d1) at (0,0,0){};
        \coordinate  (d2) at (0,1,0){};
        \coordinate  (d3) at (0,1,1){};
        \coordinate  (d4) at (0,0,1){};
        \coordinate  (d5) at (0,0.5,0){};
        \coordinate  (d6) at (0,.5,1){};
        \coordinate  (d7) at (0,0,.5){};
        \coordinate  (d8) at (0,1,.5){};
        \draw[dash pattern={on 2pt off 3pt}, black, line width=0.01mm] (d2) -- (d3) -- (d4);
        \draw[dash pattern={on 2pt off 3pt}, black, line width=0.01mm] (d5) -- (d6);
        \draw[dash pattern={on 2pt off 3pt}, black, line width=0.01mm] (d7) -- (d8);
        
        \coordinate  (d11) at (.5,0,0){};
        \coordinate  (d12) at (.5,1,0){};
        \coordinate  (d13) at (.5,1,1){};
        \coordinate  (d14) at (.5,0,1){};
        \coordinate  (d15) at (.5,0.5,0){};
        \coordinate  (d16) at (.5,.5,1){};
        \coordinate  (d17) at (.5,0,.5){};
        \coordinate  (d18) at (.5,1,.5){};
        \draw[dash pattern={on 2pt off 3pt}, black, line width=0.01mm] (d11) -- (d12) -- (d13) -- (d14) -- cycle;
        \draw[dash pattern={on 2pt off 3pt}, black, line width=0.01mm] (d15) -- (d16);
        \draw[dash pattern={on 2pt off 3pt}, black, line width=0.01mm] (d17) -- (d18);

        \coordinate  (d21) at (1,0,0){};
        \coordinate  (d22) at (1,1,0){};
        \coordinate  (d23) at (1,1,1){};
        \coordinate  (d24) at (1,0,1){};
        \coordinate  (d25) at (1,0.5,0){};
        \coordinate  (d26) at (1,.5,1){};
        \coordinate  (d27) at (1,0,.5){};
        \coordinate  (d28) at (1,1,.5){};
        \draw[dash pattern={on 2pt off 3pt}, black, line width=0.01mm] (d21) -- (d22) -- (d23) -- (d24) -- cycle;
        \draw[dash pattern={on 2pt off 3pt}, black, line width=0.01mm] (d25) -- (d26);
        \draw[dash pattern={on 2pt off 3pt}, black, line width=0.01mm] (d27) -- (d28);  
        
        \draw[dash pattern={on 2pt off 3pt}, black, line width=0.01mm] (d2) -- (d22);
        \draw[dash pattern={on 2pt off 3pt}, black, line width=0.01mm] (d3) -- (d23);
        \draw[dash pattern={on 2pt off 3pt}, black, line width=0.01mm] (d4) -- (d24);
        \draw[dash pattern={on 2pt off 3pt}, black, line width=0.01mm] (d5) -- (d25);
        \draw[dash pattern={on 2pt off 3pt}, black, line width=0.01mm] (d6) -- (d26);
        \draw[dash pattern={on 2pt off 3pt}, black, line width=0.01mm] (d7) -- (d27);
        \draw[dash pattern={on 2pt off 3pt}, black, line width=0.01mm] (d8) -- (d28);
        
        \draw[thick,->] (0,0,0) -- (1.2,0,0) node[anchor=north east]{\small $f_1 = ``cat"$};
        \draw[thick,->] (0,0,0) -- (0,1.1,0) node[anchor=north west]{\small $f_2 = ``sad"$};
        \draw[thick,->] (0,0,0) -- (0,0,1.1) node[anchor=south]{\small $f_3 = ``happy"$};

        \coordinate  (q1) at (.5,0,.5){};
        \coordinate  (q2) at (.5,0,0){};
        \coordinate  (q3) at (.5,.5,0){};
        \coordinate  (q4) at (.5,1,.5){};
        \shade[ball color = black!50, opacity =  1] (q1) circle (.025cm);
        \shade[ball color = black!50, opacity =  1] (q2) circle (.025cm);
        \shade[ball color = black!90, opacity =  1] (q3) circle (.025cm);
        \shade[ball color = black!90, opacity =  1] (q4) circle (.025cm);
        \node[above left,black!80] at (q1) {\small $d_1$};
        \node[above left,black!80] at (q2) {\small $d_2$};
        \node[below right,black!100] at (q3) {\small $d_3$};
        \node[below right,black!100] at (q4) {\small $d_4$};
        \node[text width=10cm,align=left] at (0,.76,-.9) {\small $d_1 =$ ``The \textbf{cat} makes me \textbf{happy}."- POS\\$d_2 =$ ``The \textbf{cat} is beautiful."- POS\\$d_3 =$ ``The \textbf{cat} makes me \textbf{sad}."- NEG\\$d_4 =$ ``I was \textbf{happy}, the \textbf{sad cat} made me \textbf{sad}."- NEG\\  } ;

    \end{tikzpicture}
    \caption{Classified feature space}
    \label{fig:featurespace}
\end{figure}

Let us give some brief background information on how to perform SC, mainly to clarify design choices made in the experimental set up. The pipeline for SC can be segmented into three parts: data preprocessing, document representation and classification. The next paragraphs describe the pipeline. For a lower dimensional visualization, see Figure \ref{fig:featurespace}.

\subsection{Preprocessing}
In the preprocessing part one tries to remove noise from the text that does not hold sentiment information in order to reduce complexity, i.e. dimensionality. Common techniques to do so are stop word removal, lower casing words, spelling correction and removing punctuation. Other techniques, including part of speech tagging, stemming and lemmatization, attempt to find implicit sentiment information by predicting syntactics of words or combinations of words.

\subsection{Document representation}
The second part of the pipeline for sentiment classification deals with representing documents in a mathematical interpretable fashion. The most upfront approach is to use words as features, and word counts as feature values without looking at word order, a bag-of-words approach. With $n_F$ words or combinations of words as features, $n_F=|\mathcal{F}|$, we can build an $n_F$-dimensional feature space in which each word is represented by an $n_F$-dimensional vector. The number of occurrences of a feature in the document determines the value in that dimension. Each document is then represented as the sum of all the vectors of its features. When using $n_F$ features, a corpus of $n_d$ documents can be represented as a matrix $X \in \mathbb{N}^{n_d \times n_F}$ which is referred to as the feature value matrix. $X$ represents a projection of the empirical distribution $\mathbb{P}$. In addition, one could choose to use combinations of $N$ words as features, called $N$-grams. Often, features are weighted to assign less value to more common words such as stop words. A widely used weighting scheme borrowed from information retrieval systems is the TF-IDF weighting scheme. To reduce dimensionality, feature selection algorithms can be used, selecting the most class label informative features by using for example $\chi^2$, mutual information or lexicon based selection \cite{khan2016swims}. Another popular approach for document representation uses word embeddings \cite{le2014distributed} \cite{mikolov2013efficient}, a lower dimensional projection of a high dimensional feature space. To construct a good projection, we need a lot of data, preferably from the domain that is under evaluation. 

\subsection{Classification}
When we have a mathematical representation for the source and target documents, we try to find a hypothesis function that separates the feature space in subspaces belonging to the different class labels. All documents that are represented in  certain subspace, are predicted to have the accompanying label of that specific subspace. The hypothesis function, or classifier, is constructed by minimizing a loss function, see \eqref{eqn:hoptimal_loss}. Popular classifiers for sentiment classification are Support Vector Machines, Naive Bayes and Logistic Regression (LR).

\section{Experimental Setup} \label{ex_setup}
In this section we briefly elaborate on different datasets we used for our experiments as well as the design choices we consider concerning the SC pipeline and the proposed distance measures.


\subsection{Datasets} \label{datasets}
To evaluate the methodology, we choose two sets of data that are different in their nature of topic variety, source medium, number of documents, document length, and class balance. Our primary goal to conduct two experiments with datasets at different homogeneity levels is to investigate the performance and robustness of our proposed method with respect to the diversity of the training datasets.

\subsubsection{Homogeneous dataset}
The homogeneous dataset consisting of 20 corpora from the DRANZIERA benchmark dataset \cite{dragoni2016dranziera}. We use 5000 positive and 5000 negative documents from each corpus. The documents range from 64 to 123 words with an average of 90. Each corpus consists of Amazon reviews on products from a different category. This set is referred to as the ``homogeneous" set as the nature of all corpora are similar, but only the subtopics of the corpora differ.

\subsubsection{Heterogeneous dataset}
Our second dataset contains 13 corpora. Each corpus consists of a different number of documents ranging from 502 to 30602 documents with an average of 6326 documents. The average length of each document among the corpora varies between 10 words and 20 words with an average of 15 words. We deploy a corpus of online reviews on movies, venues and consumer products \cite{kotzias2015group}, short opinions about movies \cite{Michigan2017sentiment}, and nine sets of tweets with various topics \cite{CrowdFlower2017data}. The acquired datasets do not have any major class imbalances. This set is referred to as the ``heterogeneous" set as the nature of each corpus differs in number of documents, source and topic. In a direct comparison with the first datasets, we note that the homogeneous set is richer and more balanced in terms of data points, and the topics of the corpora lay closer to each other in semantic meaning.

\subsubsection{Adaptive learning dataset}
The third set acquired is a well known dataset containing Amazon reviews on four different categories of products \cite{blitzer2007biographies}. We use 2000 reviews from each category. Motivated by the original paper \cite{ganin2016domain} on DANN, This dataset is used in order to evaluate the performance of our proposed method in an adaptive learning setting.


\subsection{Design choices}
Considering the SC pipeline design choices, in our experiments we only remove punctuation and lower case all words as prepossessing. For classification in our experiments, we use a bag-of-words approach with unigrams and bigrams and use all features that occur more than 4 times in the training corpus and at maximum in 40\% of the training documents. We weight the features according to the TF-IDF weighting scheme. We deliberately do not use word embeddings since the lower dimensional projection needs too much textual domain data to construct. Using data from other domains will distort results as the representation becomes dependent on other domains than the source and target domain that are under performance evaluation. We use LR for its good performance. We use binary classification, $y \in \{0,1\}$, giving the following loss function for our classifier

\begin{equation*}
\label{eqn:loss_functions}
\ell(x,y)_{LR} = \sum\limits_{i=1}^{n_d} \log \big( 1 + e^{-(2y_i-1) x_i \beta + \beta_0} \big) + \alpha \frac{1}{2}\|\beta\|^2 
\end{equation*}

\noindent where $w$ and $w_0$ represent the optimizers, $\alpha$ represents the regularization parameter and the hypothesis function is defined as 

\begin{equation*}
\label{eqn:hyp_func}
h(x)=\begin{cases}
        \begin{aligned}
               1 & ~ \text{if} ~ \beta x + \beta_0 > 0\\
               0 & ~ \text{else}.
        \end{aligned}
        \end{cases}
\end{equation*}

\noindent For $\alpha$, we use the default classifier settings of our used toolbox \cite{scikit-learn}.

In the adaptive learning setting, we use exactly the same model settings as in \cite{ganin2016domain}. The model deploys DANN for classification both with an original representation of 5000 features as with a marginalized Stacked Denoising Auto-encoders (mSDE) representation \cite{chen2012marginalized}, an advanced representation that improves performance in SC. For both representations, we evaluate the performance of our methodology.

For all distance measures, we only measure the distance over the $n=1000$ most occurring features for computational convenience. We do not compare the distribution of all features for two reasons. In the first place, the distance calculations will take long when using all features. And second, and more important, distributions of rarely occurring features are less reliable. 
For the Chi2 divergence, we smoothen the distributions by adding a constant $\lambda = 0.05$ over the entire distribution $p(X)$ to prevent the distance between two discrete distributions to become infinite when we encounter a probability of zero. For the KLD we add a constant $\lambda = 10^{-5}$ to the entire distribution for same reasons. For the MMD we choose a Reproducing Kernel Hilbert Space that corresponds to the Gaussian Radial Basis Function kernel $K(x,y)=e^{\frac{||x-y||^2}{2\sigma^2}}$ where $\sigma$ is our optimizer. We use corpora of the same number of objects, i.e. documents, as input by using a subset of the largest corpus of the two corpora used. For computational convenience we use a maximum of 5000 documents. For the EMD we use the set of all continuous Lipschitz functions with respect to the 1-norm. Note that the distribution is highly dependent on the length of a document. Therefore, we match document lengths among the two domains before calculating the EMD. For the optimization for the two IPMs we use available code for discrete distributions \cite{Asch2012mmd} \cite{Mayner2017pyemd} \cite{pele2008} \cite{pele2009}.

For the inner domain classification error that is used as predictor in the CMEK model, we use the same logistic regression as for the cross domain classification and use 10-fold cross validation to assess the inner domain classification error. 

\section{Results} \label{results}

In this section we will objectively present the results of our experiments according to our performance evaluation described in subsection \ref{evaluation}. 

\begin{figure}[h]	
	\centering
	\begin{subfigure}[t]{2.5in}
		\centering
		\includegraphics[width=2.5in]{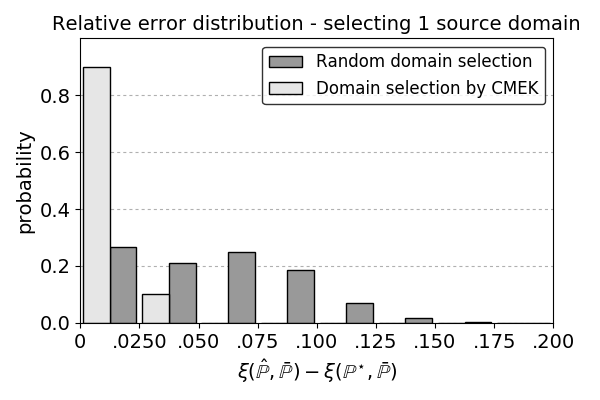}
		\caption{Homogeneous dataset}\label{fig:3a}		
	\end{subfigure}
	\quad
	\begin{subfigure}[t]{2.5in}
		\centering
		\includegraphics[width=2.5in]{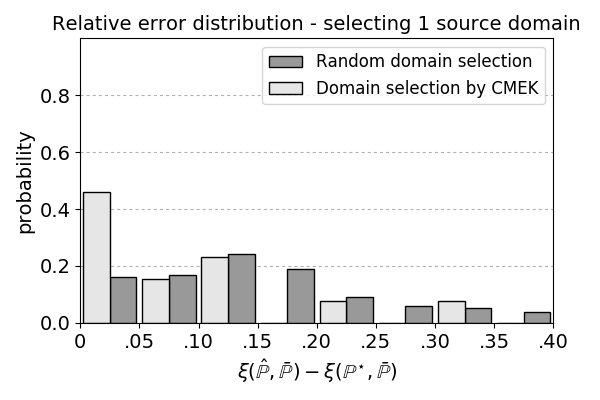}
		\caption{Heterogeneous dataset}\label{fig:3b}
	\end{subfigure}
	\caption{Relative error of CMEK and random source domain selection when selecting a single training domain}
	\label{fig:fig3}
\end{figure}

Figure \ref{fig:fig3} shows the empirical distribution of the relative error as defined in \eqref{eqn:relative_error} when selecting one source domain. The relative error distributions when randomly selecting a source domain and when using the CMEK source domain selection model are shown.

 For the homogeneous and heterogeneous dataset, the CMEK model uses the optimized weights 

\begin{equation} \label{eqn:beta}
\begin{split}
\hat{\beta}_{homogeneous} &= [0.13, ~0.52, ~1.02, ~0.00, ~0.00, ~0.11] \\
\hat{\beta}_{heterogeneous} &= [1.13, ~0.13, ~0.07, ~0.00, ~0.45, ~0.27] 
\end{split}
\end{equation}

\noindent for respectively, Chi2, MMD, EMD, KLD, the source domain inner domain classification error, and the constant 1. 

Table \ref{table:results_table} shows the probability of selecting the true best domain, the probability of selecting one of the five worst domains in terms of the cross domain classification error, and the average absolute cross domain classification error, averaged over the all target domains. We list the results for using the CMEK model for domain selection, when the source domain is selected at random and an optimal model that would always select the best source domain. The optimal cross domain classification error can be seen as lower bound of error. For the homogeneous dataset, the CMEK model uses the optimized weights

\begin{table}
\caption{Performance of CMEK, random and optimal selection.}
\label{table:results_table}

\begin{subtable}{\linewidth}
\small
\renewcommand{\arraystretch}{1.3}
\centering
\begin{tabular}{|c||c|c|c|}
\hline
                   & \textbf{Probability} & \textbf{Probability}  &  \textbf{Average}\\
\textbf{Selection} & \textbf{Best possible} & \textbf{One of 5 worst}  & \textbf{absolute} \\
\textbf{method} & \textbf{domain selected} & \textbf{domains selected} & $\mathbf{\xi(\hat{\mathbb{P}},\bar{\mathbb{P}})}$\\
\hhline{|=#=|=|=|}
Optimal & 1 & 0 & .139\\
\hline
CMEK & .600 & 0 & .144\\
\hline
Random & .053 & .263 & .191\\
\hline
\end{tabular}
\caption{Homogeneous dataset}
\end{subtable}
\begin{subtable}{\linewidth}
\small
\renewcommand{\arraystretch}{1.3}
\centering
\begin{tabular}{|c||c|c|c|}
\hline
                   & \textbf{Probability} & \textbf{Probability}  &  \textbf{Average}\\
\textbf{Selection} & \textbf{Best possible} & \textbf{One of 5 worst}  & \textbf{absolute} \\
\textbf{method} & \textbf{domain selected} & \textbf{domains selected} & $\mathbf{\xi(\hat{\mathbb{P}},\bar{\mathbb{P}})}$\\
\hhline{|=#=|=|=|}
Optimal & 1 & 0 & .252\\
\hline
CMEK & .385 & .154 & .337\\
\hline
Random & .083 & .417 & .403\\
\hline
\end{tabular}
\caption{Heterogeneous dataset}
\end{subtable}

\end{table}

\begin{figure}	
	\centering
	\begin{subfigure}[t]{2.5in}
		\centering
		\includegraphics[width=2.5in]{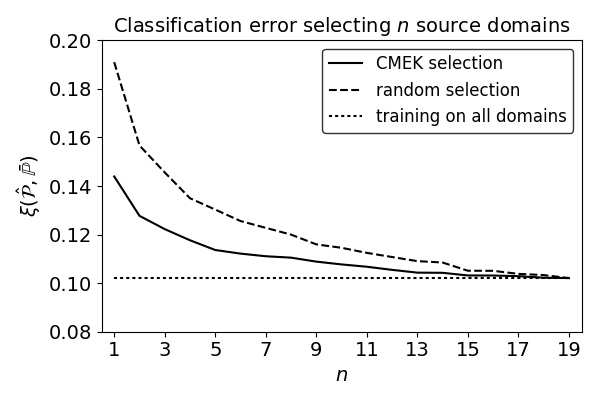}
		\caption{Homogeneous dataset}\label{fig:4a}		
	\end{subfigure}
	\quad
	\begin{subfigure}[t]{2.5in}
		\centering
		\includegraphics[width=2.5in]{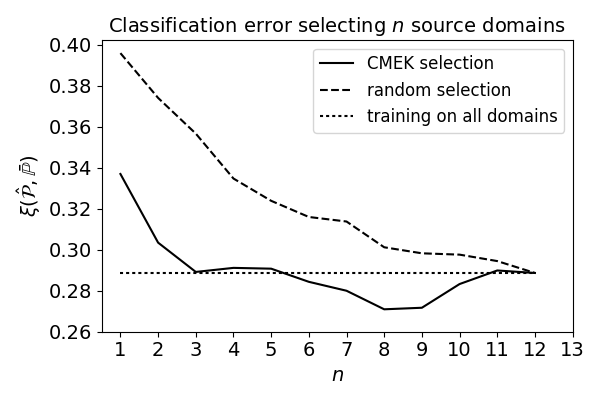}
		\caption{Heterogeneous dataset}\label{fig:4b}
	\end{subfigure}
	\caption{Performance of the CMEK model and random selection for different number of selected source domains in comparison with training on all source domains}
	\label{fig:fig4}
\end{figure}

Figure \ref{fig:fig4} shows the absolute cross domain classification error when we let our CMEK selection model select $n$ source domains. Note that the CMEK model is constructed by optimizing the predicted cross domain classification error when training the classifier on single source domains, not training on multiple. The results are compared with randomly selecting $n$ domains and training on all candidate source domains. 

\begin{table}[ht]
\caption{Performance of CMEK, random, and optimal selection in adaptive learning setting \cite{ganin2016domain}.}
\label{table:results_table2}

\begin{subtable}{0.5\linewidth}
\scriptsize
\renewcommand{\arraystretch}{1.3}
\centering
\begin{tabular}{|c||c|c|}
\hline
                   & \textbf{Probability}   &  \textbf{Average}\\
\textbf{Selection} & \textbf{Best possible}  & \textbf{absolute} \\
\textbf{method} & \textbf{domain selected}  & $\mathbf{\xi(\hat{\mathbb{P}},\bar{\mathbb{P}})}$\\
\hhline{|=#=|=|}
Optimal & 1  & .199\\
\hline
CMEK & 1  & .199\\
\hline
Random & .33  & .237\\
\hline
\end{tabular}
\caption{Original representation}
\end{subtable}
\hspace{-.5cm}
\begin{subtable}{0.5\linewidth}
\scriptsize
\renewcommand{\arraystretch}{1.3}
\centering
\begin{tabular}{|c||c|c|}
\hline
                   & \textbf{Probability}   &  \textbf{Average}\\
\textbf{Selection} & \textbf{Best possible}   & \textbf{absolute} \\
\textbf{method} & \textbf{domain selected}  & $\mathbf{\xi(\hat{\mathbb{P}},\bar{\mathbb{P}})}$\\
\hhline{|=#=|=|}
Optimal & 1  & .152\\
\hline
CMEK & 1  & .152\\
\hline
Random & .33  & .187\\
\hline
\end{tabular}
\caption{mSDA representation}
\end{subtable}

\end{table}

Table \ref{table:results_table2} reports the probability of selecting the best domain in terms of the cross domain classification error, and the respective average absolute cross domain classification error; the result is empirically computed based on the four available target domains provided by \cite{blitzer2007biographies}. We list the results corresponding to the CMEK model as a means to select the source domain, when the source domain is selected at random and an optimal model that would always select the best source domain.

\section{Conclusion} \label{conclusion}
To select a suitable source domain in terms of cross domain classification error, we hypothesized that this error can be predicted by a function of the source domain distribution $\mathbb{P}$ and marginal target domain distribution $\bar{\mathbb{P}}_x$. We proposed the CMEK model to select the best source domain(s) from a set of candidates. For our case study, the CMEK model uses a linear combination of the Chi2, MMD, EMD, KLD, the inner domain classification error and the constant 1, with weight vector $\beta$ as predictor for the cross domain classification error. To evaluate performance, we consider an example including $N-1$ distinct domains forming $N(N-1)$ different source-target domain pairs. The proposed metric parameter $\beta$ is optimized to minimize the absolute error of the prediction. The optimized predictor was used to select one or multiple best source domains among those $N-1$ domains in order to train a classifier for the $Nth$, unseen target domain. This classifier is tested by calculating the cross domain classification error of the selected source domain(s) and the target domain. The process is repeated $N$ times, each time using a different unseen target domain to test on. We benchmark the performance of our solution with randomly selecting a source domain as well as training on all candidate source domains. We perform this method over a homogeneous dataset ($N=20$) and for a heterogeneous dataset ($N=13$).

From figure \ref{fig:fig3} we see that the CMEK model is well able to identify source domains with low cross domain classification error. In 90\% of the runs on the homogeneous datasets, the selected source domain was within 2.5 percent points error of the optimal choice whereas random selection only selected 26\% of the times a source domain in this category. For the heterogeneous set we see a similar increase in the performance of the CMEK model when looking at the probability a domain is selected within 5 percent points error of the optimal choice, 45\%, and 16\%, respectively.

From Table \ref{table:results_table}, we see that compared to random domain selection, the CMEK realizes a significant improvement in average cross domain classification error for both the homogeneous set and the heterogeneous set ($p_{hom}=3.6 \times 10^{-9} ,p_{het}=0.0071$). The CMEK model has a significant larger probability of selecting the best domain compared to random selection ($p_{hom} = 3.8 \times 10^{-11}, p_{het}=0.0029$). If we look at the probability of not selecting one of the five worst domains for training, the CMEK model performs significantly better than the random selection model ($p_{hom}=2.1 \times 10^{-9}, p_{het}=0.045$). 

From \eqref{eqn:beta} we see that the CMEK model gives different weights to the distance measures for the two used datasets. This suggests that some of the considered measures are relatively better in predicting the cross domain classification error in homogeneous datasets, and the others are better when using heterogeneous datasets. The CMEK model optimizes the weights accordingly.

Although the proposed CMEK selection model significantly improves performance compared to random selecting one source domain, we would still be better off training on all the available source domains. However, if we let our model select multiple source domains to train on, we are able to get better performance than training on all source domains for some $n$, see Figure \ref{fig:4b}. For the heterogeneous dataset, the CMEK model seems to perform significantly better than training on all domains when it selects 9 or 10 domains ($p_{het}=0.043$ and $p_{het}=0.043$). Note, for all mentioned significant results, the normality assumption of the t-test is confirmed, however in this specific evaluation, the pairwise differences seem not normally distributed for $n=10$. What the optimal number of domains to train on is, may be very dependent on what candidate source domains are available. When we have candidate source domains that are somehow similar to the target domain, the optimal $n$ will be higher, or even equal to the total number of sets, as in the heterogeneous dataset, see Figure \ref{fig:4a}. For a diverse set of candidate source domains the optimal $n$ will be lower.

Looking at the results when combining the CMEK model with adaptation learning, Table \ref{table:results_table2}, we observe that the CMEK model enabled us to select the best source domain out of three candidates for each of the four target domains. This gives us a significant improvement with regard to a random selection model ($p=0.012$). 

In the light of general performance, we would recommend the CMEK model to select a source domain for an SC task for an unlabeled target domain. The CMEK model shows significantly good performance and stable behavior in selecting multiple source domains and it has solid performance in selecting the single best domain for both our homogeneous and heterogeneous datasets. We notice that the CMEK model also adds value in the evaluated adaptive learning setting.

In retrospect to our hypothesis \eqref{eqn:hypothesis}, we can conclude that it is to some extend possible to approximate the cross domain classification error as we were able to use this approximation for successful source domain selection. However, there is much room for improvement. 

\subsection{Further discussion} 

Reflecting on our conclusion, we would like to place some remarks. We showed that the CMEK model works superior in selecting one source domain compared to random selection. However, the results show that, even for candidate source domains with a wide spread in topic and source medium, it is quite beneficial to train on all the candidate source domains. One of the reasons not to train on all data could be that it is too computational expensive. Another reason might be that the candidate source domains are too diverse. To establish if this is the case, we would need a measure that informs us about the diversity of the candidate source domains. 

When we know we have candidate source domains that are similar to each other in terms of expressing sentiment, it might very well be that our CMEK model is not able to improve performance compared to training on all data. We illustrated this by using the homogeneous dataset, see Figure \ref{fig:4a}. Therefore, if we have candidate source domains that are somewhat similar to each other, we may choose to train on all domains.

Another disadvantage of the CMEK model, is that it uses some distance measures that are quite expensive to calculate. This can be a problem when we have many or large candidate source domains. This challenge might be addressed by using less features and more informative features to calculate distance over and we could optimize used code on efficiency. We might investigate what happens to the performance of the CMEK model in case we leave out the EMD and KLD measure, as they are given fewest importance, i.e. lowest weight $\beta_i$.

We constructed, tested, and evaluated the model on three different acquired datasets. We would like to remind that sentiment expression is characterized by a severe notion of stochasticity. It is hard to tell how much performance of the CMEK model will deviate when other sets are chosen to construct the model. For more available domains to construct the model, we may assume that the model will be more refined and better performing due to the wider support for the $\beta$ weight vector. 

The dataset used for the adaptive learning setting is quite popular in the field. However, it only contains data from four domains. When using one of the domains as the target, only three domains remain in the candidate source domain set. The chance of coincidentally selecting the optimal domain is not negligible. Although we presented that the results are statistically significant, for future work, we would prefer to use a dataset that contains data originated from more domains.

Although we described a method that can be easily implemented in other fields of machine learning that encounter a domain shift such as computer vision, fraud detection or spam detection, we only showed our model works reasonably well in the domain of SC.

\subsection{Future Work}
As the CMEK model is to some extension able to select a source domain with lower than average cross domain classification error, we concluded that the used measures are to some extend able to predict the cross domain classification error in order to select a suitable source domain. However, if our model would fully do justice to our hypothesis \eqref{eqn:hypothesis}, we would improve even further. We could say that proof of concept is given: it is possible to roughly predict the cross domain classification error based on the distribution of the source domain distribution and marginal distribution of the target domain. However, we still see much room for improvement.  As the lower bound for the error when selecting one domain in the heterogeneous dataset is .252, we are only half way. For the homogeneous dataset, it turns out that we are closer to achieve an optimal model.

If we would like to improve results further, first steps would be to look closer into the features over which we measure distance. Should we use the 1000 most common, or is distance better measured with more or less features? This could be dependent on corpus size as well. There might be feature selection methods to select the features that are most informative in terms of distance, instead of simply using the $N$ most occurring features. We, could for example, look at domain specific emotion lexicons to measure distance \cite{bandhakavi2017lexicon}. 

To improve, we might add some predictors to the linear combination, such as the document length distribution or the length of a corpus. Brief statements are on average more similar with each other than brief statements compared with long expressions. Also, if we have more objects in a source domain, we might benefit more when training on that large source domain than training on a corpus with only a few documents. However, with more elements in $\beta$, we might need more data to prevent over fitting.

We can also extend our model to give a weight factor on every source domain: source domains that are predicted not suitable get a low weight, domains with low distance to the target are given more weight. This approach leans towards importance sampling. 

We constructed a predictor for the cross domain classification error of one source domain, and used the model to select multiple source domains. If this is what we are interested in, it would make sense to calculate distances between a set of multiple domains and the target domain and make a selection model optimized on predicting the cross domain classification error when training on multiple domains. In that case, synergies of datasets will be taken into account. This approach will, however, be very computational expensive. 

Concerning the results when deploying the CMEK model in an adaptive learning setting, we evaluated the performance when selecting a source domain based on the original source data distribution. It would be interesting to see what we could achieve when we let the CMEK model select based on adapted source data distribution. In that scenario, the source data is first transformed with adaptive techniques, and then the respective distance is measured between the transformed source data and the target data in order to select the predicted best domain in an adaptive learning setting.

At last, we notice that when the available datasets are rather heterogeneous, the CMEK model is typically able to outperform the classifiers trained on all datasets by selecting an appropriate subset of the training domains. To realize this, it is essential to determine what the optimal number of domains to train on is. We could approach this question by defining an absolute threshold in distance between the source and target domain, or by clustering the candidate source domains and selecting a cluster of candidates to train on. Combining the proposed methodology with a model that selects the optimal number of domains to train on might bring a very useful solution to many machine learning problems.

\section*{Acknowledgment}
The authors are grateful to Dr. Mauro Dragoni for providing us the DRANZIERA dataset, and to the company Crunchr for support throughout the project. The third author gratefully acknowledges funding from the Swiss National Science Foundation under grant P2EZP2-165264.
	
\bibliographystyle{siam}	
\bibliography{mybibfile}


\end{document}

%% file: style_PM.tex
\makeatletter
\def\@settitle{\begin{center}%
		\baselineskip14\p@\relax
		\normalfont\LARGE\scshape\bfseries
		\@title
	\end{center}%
}
\makeatother

\makeatletter

\def\subsection{\@startsection{subsection}{2}%
	\z@{.5\linespacing\@plus.7\linespacing}{.5\linespacing}%
	{\normalfont\large\bfseries}}

\def\subsubsection{\@startsection{subsubsection}{3}%
	\z@{.5\linespacing\@plus.7\linespacing}{.5\linespacing}%
	{\normalfont\itshape}}

\usepackage[usenames, dvipsnames]{xcolor}
\definecolor{darkblue}{rgb}{0.0, 0.0, 0.45}

\usepackage[colorlinks	= true,
raiselinks	= true,
linkcolor	= darkblue, 
citecolor	= Mahogany,
urlcolor	= ForestGreen,
pdfauthor	= {Peyman Mohajerin Esfahani},
pdftitle	= {},
pdfkeywords	= {},
pdfsubject	= {},
plainpages	= false]{hyperref}

\usepackage{mathtools}
\usepackage{amsfonts}
\usepackage{amsmath}
\usepackage{bbm}
\usepackage{float}
\usepackage{tikz}
\usepackage{tikz-3dplot}
\usepackage{hhline}

\date{\today}